# Test Beams Summary

Kiyotomo Kawagoe [1*]

1 – Kobe University, Department of Physics
1-1 Rokkodai , Nada-ku, Kobe 657-8501 – Japan

ILC detectors are required to have unprecedented precision. Achieving this requires significant investment for test beam activities to complete the detector R&D needed, to test prototypes and (later) to qualify final detector system designs, including integrated system test. This document summarizes the discussion at this workshop on the test beam facilities and detector R&D programs to be performed there.

## 1 Introduction

ILC detectors are required to have unprecedented precision. Achieving this requires significant investment for test beam activities to complete the detector R&D needed, to test prototypes and (later) to qualifying final system designs, including integrated system test. To realize this efficiently, the ILC detector R&D groups and accelerator experts at test beam facilities have been working together. After the first ILC Detector Test Beam Workshop had been held at Fermilab in 2007 (IDTB07[1]), the roadmap document[2] for ILC detector R&D was written and sent to the executives of particle physics laboratories in the world. Since then the detector R&D groups, especially for the calorimetry and tracking devices, have obtained many fruitful results on their prototypes and various experiences at the test beam facilities. After LoIs of the two detector concept groups, ILD and SiD, had been validated in 2009, the detector R&D groups entered a new phase toward final system designs. For this purpose the second ILC Detector Test Beam Workshop was held at LAL Orsay in November 2009 (LCTW09[3]), and a new document is being prepared as a result of the workshop. In the test beam session of LCWS2010, which was very compact, updates since LCTW09 and future plans were reported and discussed. In addition, status of the AIDA[4] project to boost European detector R&D for accelerator experiments was presented.

## 2 Test Beam Facilities

We had three talks on test beam facilities. Some highlights of them are given in this document. For details, please refer to original contributions in these proceedings.

### 2.1 Fermilab Test Beam Facility

The Meson Test Facility (MTest) at Fermilab has been providing high energy particle beams for the R&D studies of the ILD detector. MTest gives users an opportunity to test the performance of their particle detectors in a variety of particle beams. A good example is the combined test of the CALICE collaboration, where the prototypes of electromagnetic and hadron calorimeters, as well as a muon tail catcher were combined and tested together.

---

* This work is supported by the Grant-in-Aid for Science Research, JSPS (No. 19204027).

One of the requests of the detector R&D groups is to have test beams with an ILC-like time structure (1 ms train with a 199 ms inter-train quiet period). To this purpose the Accelerator division has installed quadruple extraction hardware that can deliver beams within 1 to 5 ms short spills.

At the experimental area of MTest, new common equipments have been installed (see Fig. 1). One is the precision tracking devices: two stations of PHENIX pixel tracker (pixel size 50x400 $\mu m^2$, active area 6x6 $cm^2$) and four stations of CMS pixel tracker (pixel size 100x150 $\mu m^2$, overlap area 2x2 $cm^2$) . The other is a fast timing detector, made of quarts bar and PHOTEK MCPs, with a time resolution of 6 ps.

In addition to the MTest beam line, Fermilab is preparing a new test beam facility in the Meson Center beam line. The proposal of the new facility (MCenter Test Facility) has been approved and the facility is expected to be available after 2010 summer shutdown.

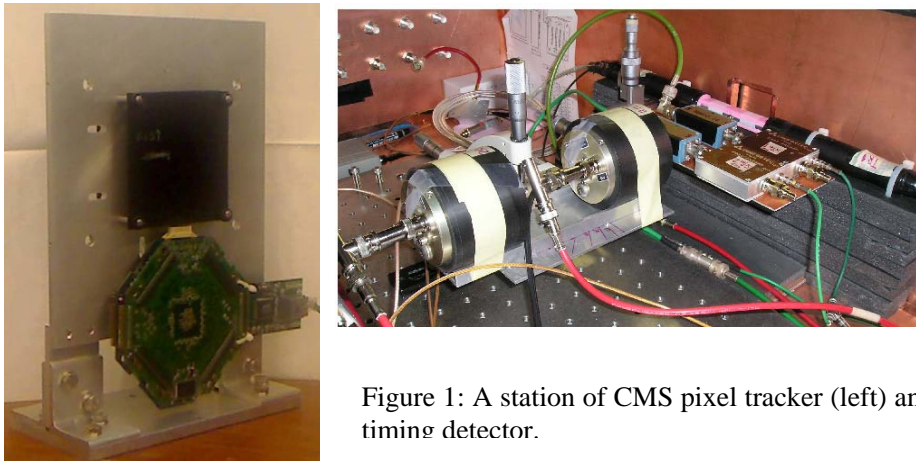

Figure 1: A station of CMS pixel tracker (left) and the fast timing detector.

### 2.2 ESTB SLAC Test Beam Project

End Station Test Beam (ESTB) is an approved and funded SLAC project to use a small fraction of the 13.6 GeV electron beam from the Linac Coherent Light Source (LCLS) to restore test beam capability in End Station A (ESA) as shown in Fig. 2. Four kicker magnets will be installed in the Beam Switch Yard to divert 5 Hz of LCLS beam to the A-line. This beam can be directed against a thin screen to produce secondary electrons or positrons with energies up to the incident energy, and a wide range of intensities including single particles per pulse suitable for detector studies. ESTB is constructed in 2010 and will be available by Spring 2011. The installation of a secondary hadron target and a hadron beam line in ESA is a possible upgrade for 2011. This beam will produce pions and kaons over a broad range of momentum, suitable detector R&D. The expected yields of secondary particles are shown in Fig. 3.

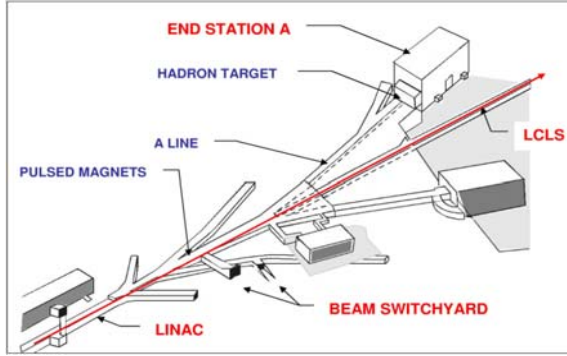

Figure 2: A schematic view of ESTB.

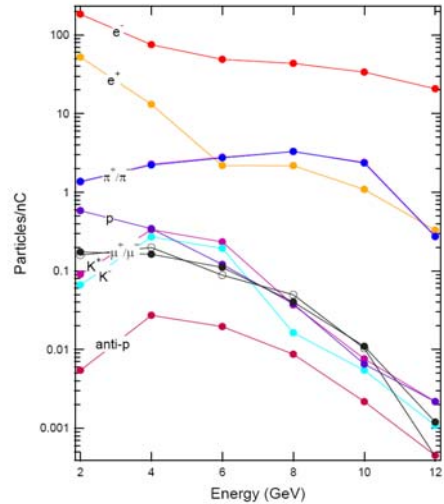

Figure 3: Expected yields of secondary particles at ESTB.

### 2.3 Asian Test Beam Facilities

There are several low energy test beam facilities in Asia, where test of small units can be performed.

#### 2.3.1 J-PARC Test Beam Facility

In Japan, the test beam line at KEK (FTBL) was shut down at the end of 2009 for the upgrade of the KEKB collider. On the other hand, a new test beam facility is being constructed in the hadron experimental hall at J-PARC (see Figure 4). The 50 GeV proton synchrotron started its operation at 30 GeV beam energy in 2009. The K1.1 Beam line will become available in fall 2010, where hadrons with momentum 0.5~1.1 GeV and good enough particle yields are available. This beam line can be used for test beam experiments until preparation of the main experiment at K1.1 is started. The K1.8BR beam line is dedicated to the test beam experiments, and hadrons with momentum 0.5~1.5 GeV are available. This beam line also will be ready in 2010. However, the particle yields are expected to be very small at the beginning to be used for the experiments, until the intensity of the proton synchrotron becomes close to the design value (100 MW).

#### 2.3.2 IHEP Beijing Test Beam Facility

In China, BTF (Beijing Test beam Facility) provides primarily electron beams with momentum 1.1~1.5 GeV and secondary beams with momentum 0.4~1.2 GeV. BTF is now under a long shut down for its upgrade. The commissioning is scheduled to start in January 2011.

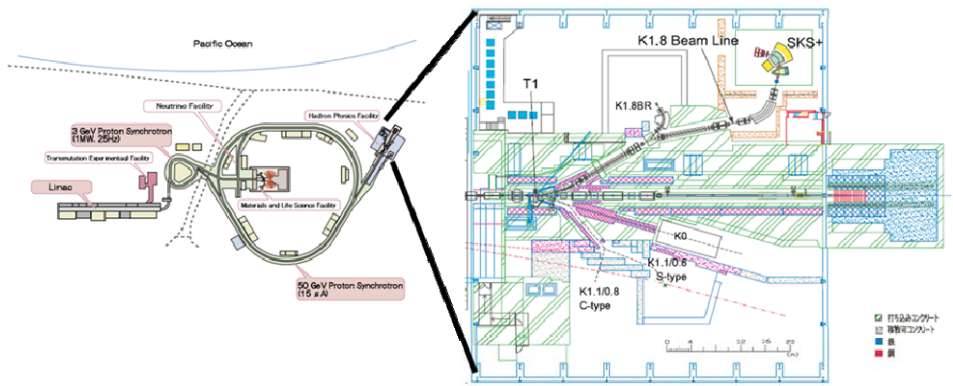

Figure 4: A schematic view of the accelerator complex at J-PARC and the test beam lines in the experimental hall.

## 3 Summary

The detector R&D groups deeply acknowledge the effort of the staffs of test beam facilities at particle physics laboratories for providing test beams in recent years. The R&D groups definitely need new or upgraded test beam facilities to continue their studies for the ILC project. To keep the momentum of the test beam activities, communication between the facility staffs and the users is essential.